\documentclass[twocolumn,superscriptaddress]{revtex4}

\usepackage{doi}
\usepackage{hyperref}
\hypersetup{
  colorlinks=true,        
  linkcolor=blue,         
  citecolor=cyan,         
}

\usepackage{graphicx}
\usepackage{dcolumn}
\usepackage{bm}
\usepackage{color}
\usepackage{enumitem}
\usepackage{amsmath}
\usepackage{amssymb}
\usepackage{orcidlink}
\begin{document}
\title{Probing the gravity of a Schwarzschild black hole in the presence of a cloud of strings with EMRIs}

\author{Mirzabek Alloqulov\orcidlink{0000-0001-5337-7117}}
\email{malloqulov@gmail.com}
\affiliation{School of Physics, Harbin Institute of Technology, Harbin 150001, People’s Republic of China}

\affiliation{University of Tashkent for Applied Sciences, Str. Gavhar 1, Tashkent 100149, Uzbekistan}

\author{Ahmadjon Abdujabbarov,\orcidlink{0000-0002-6686-3787}}
\email{ahmadjon@astrin.uz}
\affiliation{School of Physics, Harbin Institute of Technology, Harbin 150001, People’s Republic of China}
\affiliation{Tashkent State Technical University, Tashkent 100095, Uzbekistan}

\author{Bobomurat Ahmedov,\orcidlink{0000-0002-1232-610X}}
\email{ahmedov@astrin.uz}

\affiliation{Institute for Advanced Studies, New Uzbekistan University, Movarounnahr str. 1, Tashkent 100000, Uzbekistan}
\affiliation{School of Physics, Harbin Institute of Technology, Harbin 150001, People’s Republic of China}
\affiliation{Institute of Theoretical Physics, National University of Uzbekistan, Tashkent 100174, Uzbekistan}

\author{Chengxun Yuan,\orcidlink{0000-0002-2308-6703}}
\email{yuancx@hit.edu.cn}

\affiliation{School of Physics, Harbin Institute of Technology, Harbin 150001, People’s Republic of China}

\date{\today}
\begin{abstract}
Here, we explore the effect of the cloud of strings (CoS) on the gravitational waveforms of extreme mass ratio inspirals (EMRIs). The EMRI system consists of a supermassive black hole (BH) and a compact stellar mass object moving around it. We begin with studying the test particle motion around the Schwarzschild BH surrounded by a CoS by using the Lagrangian formalism. Moreover, we investigated the effect of the CoS parameter on the evolution of the semi-latus rectum and eccentricity. We then turn to the exploration of the impact of the CoS parameter on the gravitational waveforms of the EMRI system. The analysis performed shows that Laser Interferometer Space Antenna (LISA)  could detect the CoS imprint in gravitational waveforms when the values of the string cloud parameter $\alpha \gtrsim 2 \times 10^{-6}$. 
\end{abstract}

\maketitle

\section{Introduction}

In general relativity (GR), black holes (BHs) are exact solutions of Einstein's field equations. Firstly, the vacuum solutions of the field equations were obtained by Karl Schwarzschild~\cite{1916SPAW.......189S,2015arXiv151202061B} and Roy Kerr~\cite {1963PhRvL..11..237K}, which represent the spherically and axially symmetric solutions, respectively. Over the years, various solutions have been obtained by considering various theories of gravity. Now, we can classify the BHs according to their mass, which are primordial BHs, stellar-mass BHs, intermediate-mass BHs, and supermassive BHs (SMBHs). The recent direct detection of the shadows of M87*~\cite{Akiyama19L1,Akiyama19L6} and SgrA*~\cite{Event} put an end to questions about the existence of SMBHs. Extreme mass-ratio inspirals (EMRIs), which consist of a celestial body of mass $m$ such as a stellar-mass BH or neutron star orbiting a SMBH of mass $M \gg m$, are among the most promising astrophysical sources of low-frequency gravitational waves (GWs) for forthcoming space-based detectors such as Laser Interferometer Space Antenna (LISA) \cite{AmaroSeoane2018, Barack2009}, Taiji~\cite{10.1093/ptep/ptaa083}, TianQin~\cite{Gong_2021} and DECi-hertz Gravitational-wave Observatory (DECIGO)~\cite{10.1093/ptep/ptab019}. Due to their long-lasting inspiral phase and the large number of orbital cycles ($\sim 10^5$--$10^6$) emitted in the strong-gravity regime, EMRIs provide a unique probe of the geometry of spacetime in the vicinity of massive BHs. The GWs emitted from these systems carry a detailed imprint of the underlying spacetime metric, allowing precision tests of GR and the exploration of possible deviations induced by alternative theories of gravity or exotic forms of matter.

Among possible extensions of Schwarzschild geometry, the \emph{cloud of strings} (CoS) model proposed by Letelier \cite{Letelier1979} has attracted growing interest. In this model, the energy-momentum tensor of a one-dimensional string distribution modifies the Schwarzschild metric, introducing a conical deficit angle characterized by a dimensionless density parameter $\alpha$. The resulting metric retains spherical symmetry, but exhibits deviations in curvature and gravitational potential, which can significantly affect geodesic motion and, consequently, the gravitational waveforms emitted by orbiting bodies~\cite{Yazadjiev2006, Ghosh2014}. Over the years, CoS has been studied by taking into account various theories of gravity~\cite{Mustafa:2022xod,Singh:2020nwo,Ghosh:2014pga,Atamurotov:2022knb,Ganguly:2014cqa,deMToledo:2018tjq,Ashraf:2024dwg,Rincon:2018ktz,Richarte:2007bx,Li:2020zxi,Belhaj:2022kek,Ahmed:2025nnm,Ahmed:2025tbo,Al-Badawi:2025urb,Ganaie:2025nkn,Bao:2025wdp,Ahmed:2025qza}.

Investigating EMRIs in the presence of a CoS provides an effective avenue to probe the interplay between topological defects and black hole gravity. Since the orbital evolution of EMRIs is highly sensitive to small perturbations in the background metric, even weak deviations induced by the string cloud can leave measurable signatures in the GW phase and amplitude. In particular, phenomena such as zoom–whirl orbits and periastron precession can serve as direct indicators of modified spacetime geometry~\cite{Glampedakis2002, Levin2008,Wang:2025hla,Haroon:2025rzx,Jiang:2024cpe,Wang:2025wob,Lu:2025cxx,Chen:2025aqh,Zare:2025aek,Alloqulov:2025ucf,Alloqulov:2025bxh}. Detecting such signatures in future GW observations would not only offer a novel test of GR but could also constrain the possible existence and distribution of cosmic strings or related topological defects in the vicinity of SMBHs. There are different types of models to obtain gravitational waveforms of EMRIs, which are analytic kludge~\cite{PhysRev.131.435,PhysRev.136.B1224}, numerical kludge~\cite{PhysRevD.75.024005}, and adiabatic waveforms~\cite{PhysRevLett.128.231101,PhysRevD.103.104014}. Till today, various theories of gravity have been tested by EMRIs~\cite{Zhang:2024ugv,Fu:2024cfk,AbhishekChowdhuri:2023gvu,Liu:2024qci,Pani:2011xj,Yunes:2009ry,Zi:2023qfk,zi2025,ashoorioon2025,Yang:2024cnd,Xia:2025yzg}.

In this paper, we study the motion of the test particles around the Schwarzschild BH surrounded by a CoS. In addition, we investigated the impact of the CoS parameter on the dynamics of the EMRI system. In particular, the effect of the CoS parameter on the orbital evolution of the semi-latus rectum and eccentricity is studied. Moreover, we explore the gravitational waveforms of the EMRI system initially and after a period of 1 year. By comparing obtained results with the standard Schwarzschild case, we aim to quantify the influence of the string cloud on the orbital evolution and identify potential observational signatures accessible to next-generation space-based detectors.

The paper is structured as follows: we briefly review the spacetime of the Schwarzschild BH in the presence of a CoS, and we investigate the timelike geodesics by considering the Lagrangian formalism in Section~\ref{sec:spacetime}. Section~\ref{sec:fluxes} is devoted to the exploration of the orbital evolution of the EMRI system. In Section~\ref{sec:GWforms}, we study the gravitational waveforms of the EMRI system and analyze the detectability of the CoS parameter by a future space-based detector. Our conclusions and discussions are presented in Section~\ref{summary}.

\section{Spacetime and timelike geodesics}\label{sec:spacetime}

In this section, we briefly review the spacetime of the Schwarzschild BH in the presence of CoS, and investigate the timelike geodesics. The line element of the Schwarzschild BH in the presence of CoS can be written in the following form~\cite{Letelier1979}
\begin{equation}
ds^2=-f(r)dt^2+\frac{1}{f(r)}dr^2+r^2(d\theta^2+\sin^2{\theta}d\phi^2)\ , 
\end{equation}
with 
\begin{equation}
    f(r)=1-\frac{2M}{r}-\alpha\ ,
\end{equation}
where $\alpha$ refers to the cloud of string (CoS) parameter. It can be observed that we can recover the ordinary Schwarzschild BH when $\alpha=0$. It is worth noting that the spacetime is not asymptotically flat because the cloud of strings behaves as a distribution of cosmic strings that introduces a constant angular deficit, which persists even at spatial infinity.

\begin{figure*}
\centering
\includegraphics[scale=0.48]{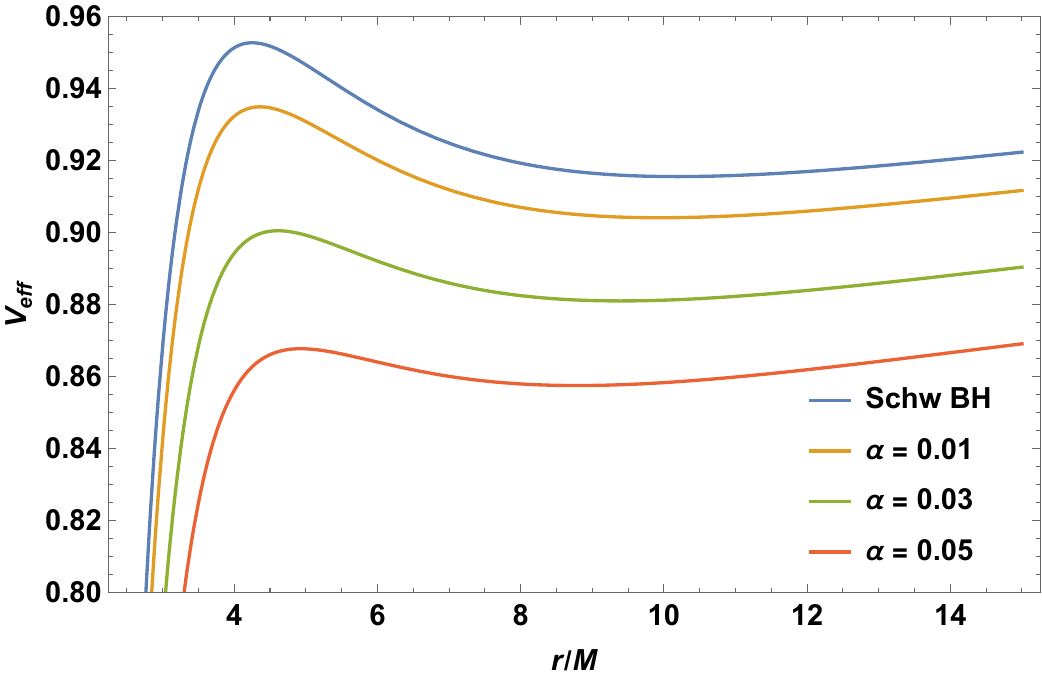}
\includegraphics[scale=0.48]{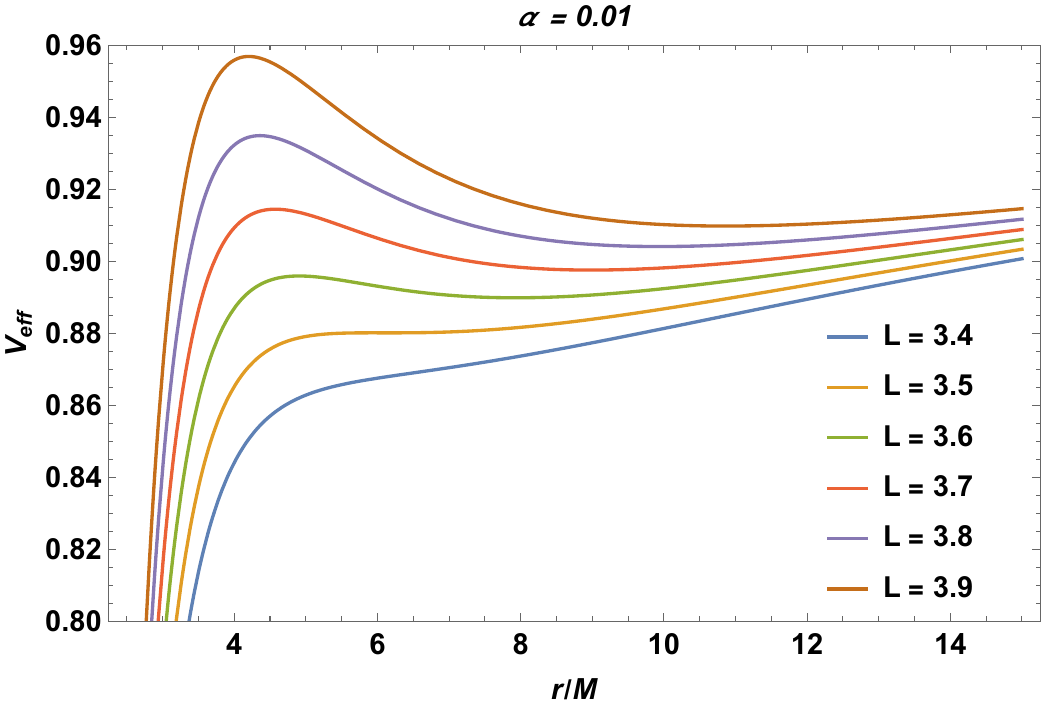}
\caption{The plot demonstrates the radial dependence of the effective potential of the massive particles for the different values of the CoS parameter (left panel) and orbital angular momentum (right panel). $\alpha=0.01$ is set for the right panel.}
\label{fig:eff}
\end{figure*}

\begin{figure*}[htbp]
\centering
\includegraphics[scale=0.48]{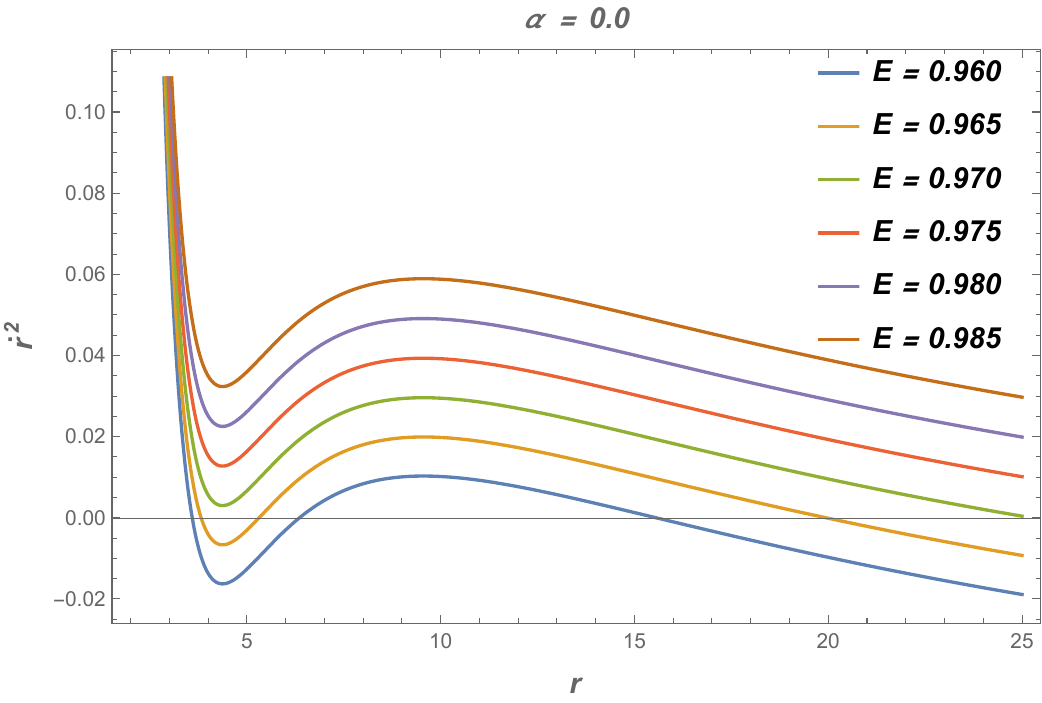}
\includegraphics[scale=0.48]{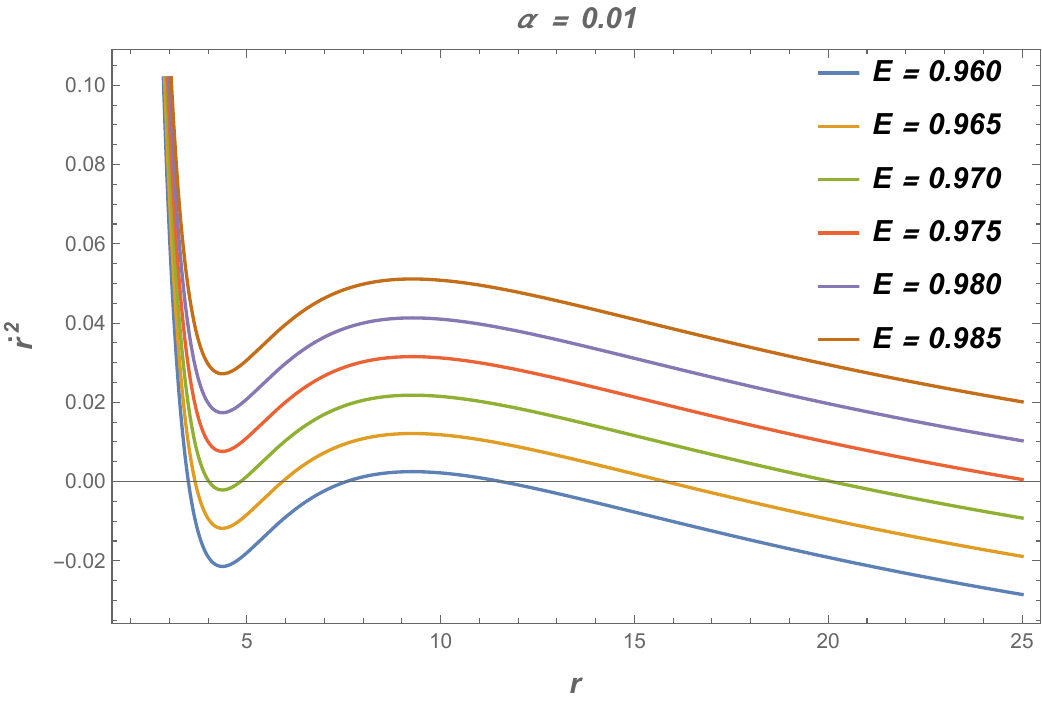}
\caption{Radial dependence of $\dot{r}^2$ for different values of the energy $E$. Here, $L$ equals to the $0.5(L_{ISCO}+L_{MBO})$.}
\label{fig:rdot}
\end{figure*}

One can write the Lagrangian for the test particle with mass $m$ in the equatorial plane ($\theta=\pi/2$) as
\begin{equation}
\mathcal{L}=\frac{1}{2}m\Big[-f(r)\dot{t}^2+\frac{\dot{r}^2}{f(r)}+r^2\dot{\phi}^2\Big]\ ,
\end{equation}
where the dot refers to the derivative with respect to the proper time. We are considering a spherical symmetric spacetime. Therefore, there are two conserved quantities, which are the energy $E$ and the angular momentum $L$. Using the Euler-Lagrange equations, we can get as follows
\begin{eqnarray}
E&=&-\frac{\partial \mathcal{L}}{\partial \dot{t}}=mf(r)\dot{t}\,,\\ \nonumber
L&=&\frac{\partial \mathcal{L}}{\partial \dot{\phi}}=mr^2\dot{\phi}\ . 
\end{eqnarray}
We can write the following expression using the normalization condition $g_{\mu\nu}u^{\mu}u^{\nu}=-1$

\begin{equation}\label{eq:eff}
\dot{r}^2+V_{eff}=E^2\,,
\end{equation}
where $V_{eff}$ refers to the effective potential for the massive particle motion, and it is given by the following equation 
\begin{equation}
V_{eff}=\Big(1-\frac{2M}{r}-\alpha\Big)\Big(1+\frac{L^2}{r^2}\Big)\,.
\end{equation}
We demonstrate the radial dependence of the effective potential of the massive particle around the Schwarzschild BH surrounded by a CoS for the different values of the CoS parameter and orbital angular momentum in Fig.~\ref{fig:eff}. It can be seen from this figure that the values of the effective potential decrease with the increase of the CoS parameter and vice versa for the orbital angular momentum. Using the effective potential, we can analyze the marginally bound orbit (MBO) and the innermost stable circular orbit (ISCO). One can write the following conditions for the MBO and ISCO
\begin{align}
V_{eff}=1\,, \quad \frac{dV_{eff}}{dr}=0\,.
\end{align}
and
\begin{align}
    \dot{r}=0\,, \quad \frac{d V_{eff}}{dr}=0 \,, \quad \frac{d^2V_{eff}}{dr^2}=0\,,
\end{align}
respectively. After that, we can numerically analyze the effect of the CoS parameter on the MBO and ISCO. 
\begin{table}[]
\centering
\resizebox{0.47\textwidth}{!}{
\begin{tabular}{cccccccccc}
\hline
\hline
$\alpha$ & $r_{MBO}$ & $L_{MBO}$ & $r_{ISCO}$ & $L_{ISCO}$ & $E_{ISCO}$  \\ \hline
 0.00 & 4.00000 & 4.00000 & 6.00000 & 3.46410 & 0.942809 \\
 0.01 & 4.00154 & 4.08082 & 6.06061 & 3.49909 & 0.938083 \\
 0.02 & 4.00594 & 4.16333 & 6.12245 & 3.53480 & 0.933333 \\
 0.03 & 4.01294 & 4.24763 & 6.18557 & 3.57124 & 0.928559 \\
 0.04 & 4.0223 & 4.33381 & 6.25000 & 3.60844 & 0.923760 \\
 0.05 & 4.03385 & 4.42198 & 6.31579 & 3.64642 & 0.918937 \\ 
\hline
\hline
\end{tabular}
}
\caption{The values of the MBO and ISCO parameters for the different values of the CoS parameter.}
\label{table1}
\end{table}
We tabulated the values of the MBO and ISCO parameters in Table~\ref{table1}. One can see from this table that all MBO and ISCO parameters, except ISCO energy $E_{ISCO}$, increase under the influence of the CoS parameter. 
Subsequently, we rewrite the Eq.~(\ref{eq:eff}) in terms of $\dot{r}^2$ as follows

\begin{equation}
\dot{r}^2=f(r)\Big[\frac{E^2}{m^2f(r)}-\frac{L^2}{m^2r^2}-1\Big]\ . 
\end{equation}
To illustrate the radial motion, we plot $\dot{r}^2$ as a function of $r$ for different values of the energy $E$ in Fig.~\ref{fig:rdot}.  Here, the angular momentum is fixed to $L=0.5(L_{\text{ISCO}}+L_{\text{MBO}})$, where $L_{\text{ISCO}}$ and $L_{\text{MBO}}$ are the angular momenta for the ISCO and MBO, respectively.

Here, we consider the eccentric motion by using the parameterization of the radial coordinate as
\begin{equation}
r(\chi)=\frac{pM}{1+e\cos{\chi}}\,,
\end{equation}
where $p$ and $e$ represent the semi-latus rectum and eccentricity, respectively. $\chi$ refers to the reparameterized radial coordinate. Using condition $\dot{r}=0$, we can find that the extrema correspond to the periapsis $r_p$ and the apoapsis $r_a$, which are the minimum and maximum radii, respectively. We can write them as follows.
\begin{align}
r_p=\frac{pM}{1+e}, \quad r_a=\frac{pM}{1-e}\,.
\end{align}
The expressions for $E$ and $L$ can be obtained by solving $\dot{r}(r_a)=\dot{r}(r_p)=0$
\begin{eqnarray}
E^2&=& \frac{m^2 \left[(p(1-\alpha)-2)^2-4 e^2\right]}{p \left(p(1-\alpha)-e^2-3\right)}\,, \\ \nonumber
L^2 &=& \frac{m^2 M^2 p^2}{p(1-\alpha)-e^2-3}\ .
\end{eqnarray}
It should be noted that the orbital motion of the test particle is characterized by two fundamental frequencies: radial $\Omega_r$ and angular $\Omega_{\phi}$. We can obtain them as follows
\begin{align}
\Omega_r=\frac{2\pi}{T_r}, \quad \Omega_{\phi}=\frac{\Delta \phi}{T_r}\ ,
\end{align}
where
\begin{eqnarray}
T_r&=&2\int_{r_{min}}^{r_{max}}dt=\int_0^{2\pi}\frac{dt}{dr}\frac{dr}{d\chi}d\chi\,, \nonumber \\
\Delta \phi&=&2\int_{r_{min}}^{r_{max}}d\phi=2\int_0^{2\pi}\frac{d\phi}{dt}\frac{dt}{dr}\frac{dr}{d\chi}d\chi\,.
\end{eqnarray}
We can get the following expressions for the fundamental frequencies for the particles around the Schwarzschild BH surrounded by a CoS
\begin{eqnarray}
 \Omega_r&=&\frac{(1-e^2)^{3/2}(1-\alpha)^{1/2}}{M}p^{-3/2}-\frac{3(1-e^2)^{5/2}}{M(1-\alpha)^{1/2}}p^{-5/2}+ \\ \nonumber
 &&+\mathcal{O}(p^{-7/2})\,, \nonumber \\
 \Omega_{\phi} &=& \frac{(1-e^2)^{3/2}}{M}p^{-3/2}+\frac{3e^2(1-e^2)^{3/2}}{M(1-\alpha)}p^{-5/2}+\\ \nonumber
 &&+\mathcal{O}(p^{-7/2})\,.
\end{eqnarray}
\begin{figure*}[htbp]
\centering
\includegraphics[scale=0.48]{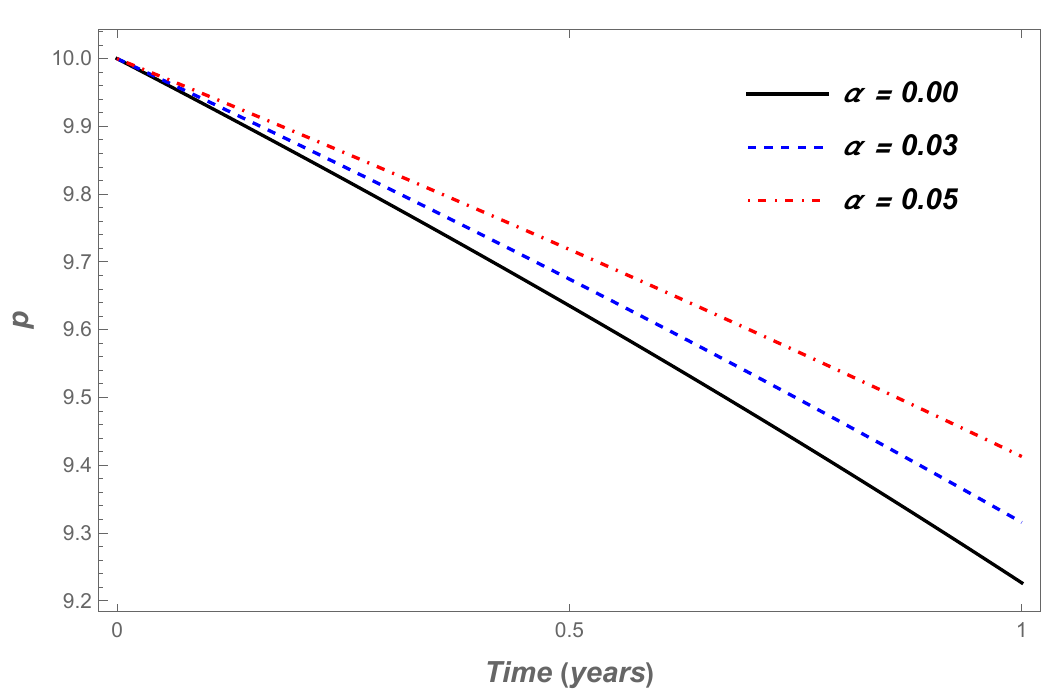}
\includegraphics[scale=0.48]{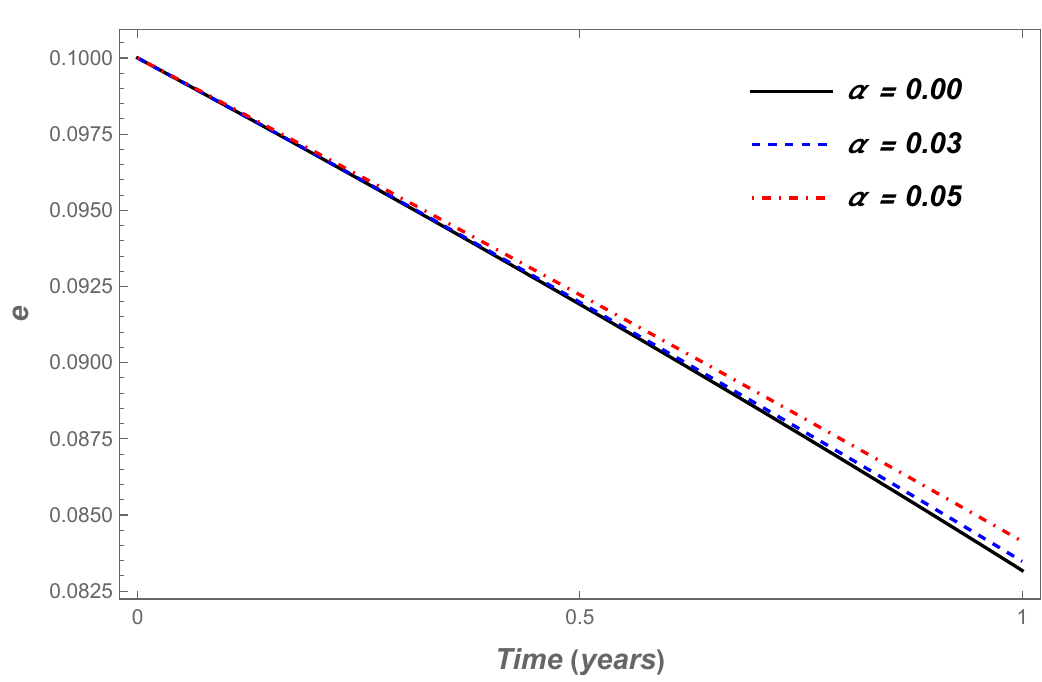}
\caption{Evolution of the semi-latus rectum $p$ (left panel) and eccentricity $e$ (right panel) for different values of the CoS parameter $\alpha$. Here, the initial values of other parameters are $(M,m,p,e)=(10^6 M_{\odot},10 M_{\odot},10M,0.1)$.}
\label{fig:pe}
\end{figure*}
{It can be observed from the above expressions that the main part of the radial frequency (which is proportional to $p^{-3/2}$) is sensitive to the existence of the cloud of strings, while the CoS parameter appears in the angular frequency in terms lower than $p^{-3/2}$. }

\section{Fluxes and orbital evolution}\label{sec:fluxes}

In this section, we consider the GW fluxes and orbital evolution from an EMRI system. An EMRI system is characterized by a compact object slowly spiralling into the SMBH, a process caused by the emission of GWs. In this paper, we employ the adiabatic approximation, where orbital parameters are held constant over one period and are only updated between cycles to reflect the accumulated effects of radiation. 
The rate of change in energy and angular momentum can be characterized by the energy flux $-dE/dt$ and the flux of angular momentum $-dL_i/dt$. They can be defined as follows~\cite{Thorne:1980ru,Maggiore:2007ulw,Xia:2025yzg}
\begin{align}
- \frac{dE}{dt} = \frac{1}{5} \langle \dddot{Q}_{ij} \dddot{Q}^{ij} \rangle\,, \quad -\frac{dL_i}{dt}=\frac{2}{5}\epsilon_{ijk}\langle\ddot{Q}^{jk}\dddot{Q}^k_m\rangle\,,
\end{align}
where $Q_{ij}$ is the quadrupole moment of the system, which equals $Q_{ij}=m(x_ix_j-\frac{1}{3}\delta_{ij}x^kx_k)$. Here, we assume that the reduced mass of the EMRI system approximately equals the mass of the compact object, i.e., $\mu\approx m$. The indices $i$, $j$ refer to the Cartesian spatial coordinates, namely $(x^1,x^2,x^3)=(r\cos{\phi},r\sin{\phi},0)$, and $\langle \cdot \rangle$ corresponds to the average over a complete period. Using the equations of motion, we can obtain the average fluxes of energy and angular momentum for the Schwarzschild BH surrounded by a CoS as follows
\begin{widetext}
\begin{eqnarray}
\left\langle \frac{dE}{dt}\right\rangle &=& \frac{(1-e^2)^{3/2}m^2}{15M^2 p^5}\Big[96+292e^2+37e^4+(372+285e^2)e^2\alpha+(60+219e^2+30e^4)e^2\alpha^2-(4+e^2)e^2\alpha^3\Big] + \nonumber \\
&+& \frac{(1-e^2)^{3/2}e^2m^2}{5M^2p^6(1-\alpha)}\Big[176+450e^2+53e^4+(-744-618e^2+278e^4)\alpha+(-224-1314e^2-335e^4+30e^6)\alpha^2+ \nonumber \\
&+& (24+42e^2+4e^4)\alpha^3\Big]+\mathcal{O}(p^{-7})\ , \nonumber \\
\left\langle \frac{dL_z}{dt}\right\rangle&=&\frac{4(1-e^2)^{3/2}m^2}{5Mp^{7/2}}\Big[8+7e^2+(14+e^2)e^2\alpha+(3+2e^2)e^2\alpha^2\Big]+ \nonumber \\
&+& \frac{4(1-e^2)^{3/2}e^2m^2}{5M(1-\alpha)}\Big[38+27e^2+(-92+12e^2+3e^4)\alpha+(-34-51e^2+6e^4)\alpha^2\Big]+\mathcal{O}(p^{-11/2})\ .
\end{eqnarray}
\end{widetext}
The standard Schwarzschild BH results are recovered when $\alpha=0$. One can use the following dependence of the total energy and angular momentum on the semi-latus rectum and eccentricity to obtain the orbital evolution of the system 
\begin{eqnarray}
\frac{dE}{dt}&=&\frac{\partial E}{\partial p}\frac{dp}{dt}+\frac{\partial E}{\partial e}\frac{d e}{dt}\,, \nonumber \\
\frac{dL_z}{dt}&=&\frac{\partial L_z}{\partial p}\frac{dp}{dt}+\frac{\partial L_z}{\partial e}\frac{de}{dt}\,.
\end{eqnarray}
Using the above relations, we can obtain equations for $\langle dp/dt \rangle$ and $\langle de/dt \rangle$ as
\begin{eqnarray}
&&\left\langle \frac{dp}{dt} \right\rangle = \frac{\frac{\partial E}{\partial e}\langle\frac{dL_z}{dt}\rangle-\langle\frac{dE}{dt}\rangle\frac{\partial L_z}{\partial e}}{\frac{\partial E}{\partial e}\frac{\partial L_z}{\partial p}-\frac{\partial L_z}{\partial e}\frac{\partial E}{\partial p}}\,, \\ \nonumber 
&&\left\langle \frac{de}{dt} \right\rangle = \frac{\langle\frac{dE}{dt}\rangle\frac{\partial L_z}{\partial p}-\langle\frac{dL_z}{dt}\rangle\frac{\partial E}{\partial p}}{\frac{\partial E}{\partial e}\frac{\partial L_z}{\partial p}-\frac{\partial L_z}{\partial e}\frac{\partial E}{\partial p}}\ .
\end{eqnarray}
We employ the numerical method to illustrate the evolution of the semi-latus rectum and eccentricity, rather than the analytical method. In Fig.~\ref{fig:pe}, the evolution of the semi-latus rectum and eccentricity was demonstrated for the different values of the CoS parameter. The black solid line corresponds to the results for the ordinary Schwarzschild BH.   
\begin{figure*}[htbp]
\includegraphics[scale=0.5]{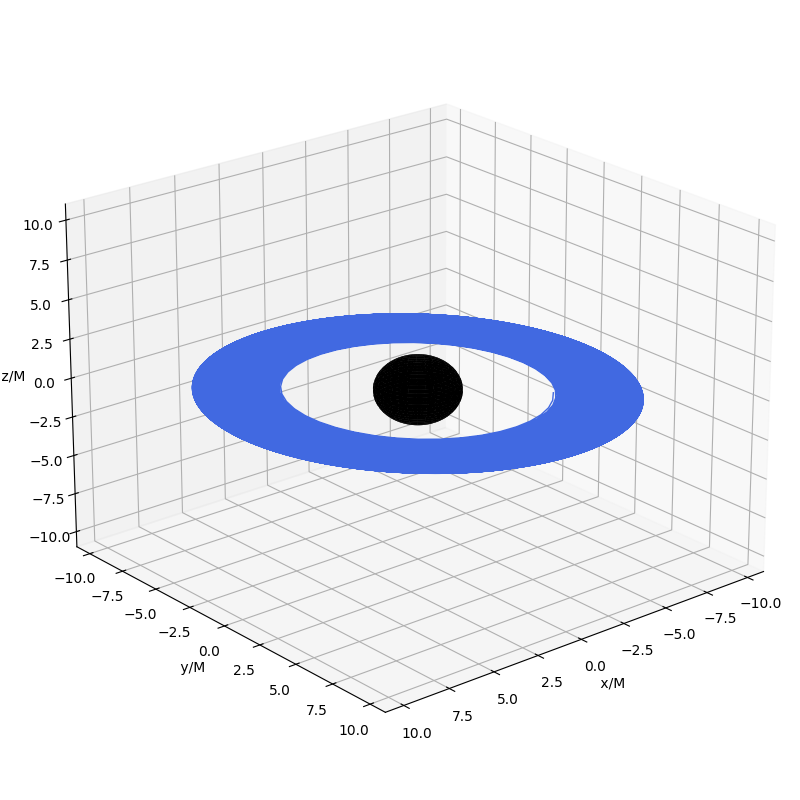}
\includegraphics[scale=0.5]{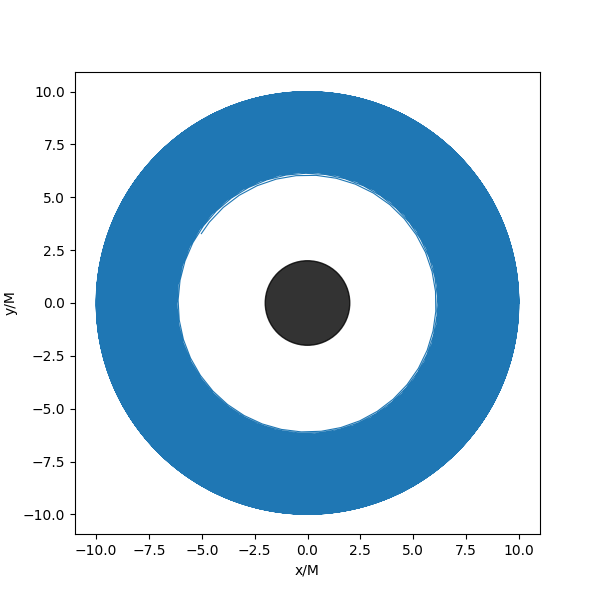}
\caption{The plot demonstrates the complete quasi-circular orbits around the supermassive Schwarzschild BH surrounded by a CoS. Here, we set the CoS parameter as $\alpha=0.01$, and $(r_0,\phi_0)=(10M,\pi/2)$.}
\label{fig:trac}
\end{figure*}
\begin{figure*}
\includegraphics[scale=0.45]{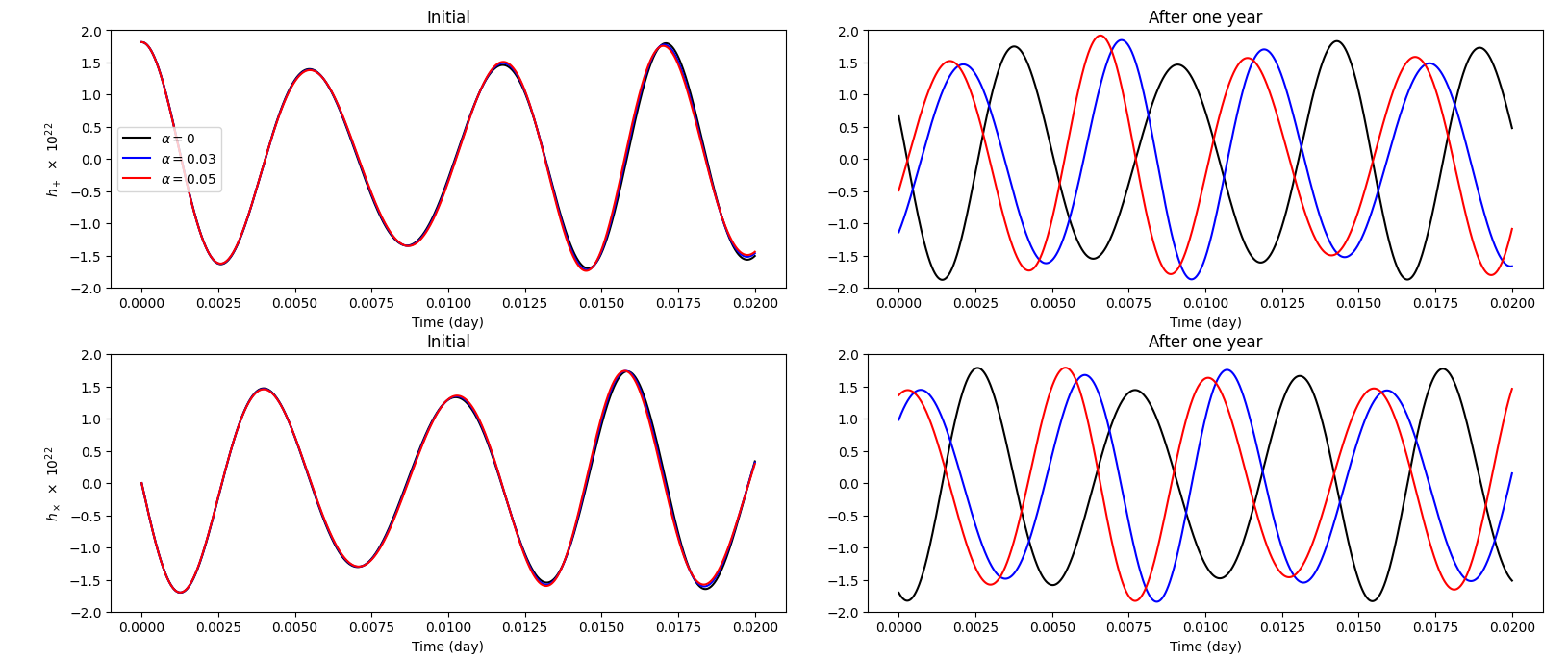}
\caption{The plot shows the gravitational waveforms of the EMRI system for the different values of the CoS parameter initially (left panels) and after a year (right panels).}
\label{fig:GW}
\end{figure*}
\begin{figure*}
    \centering
    \includegraphics[scale=0.35]{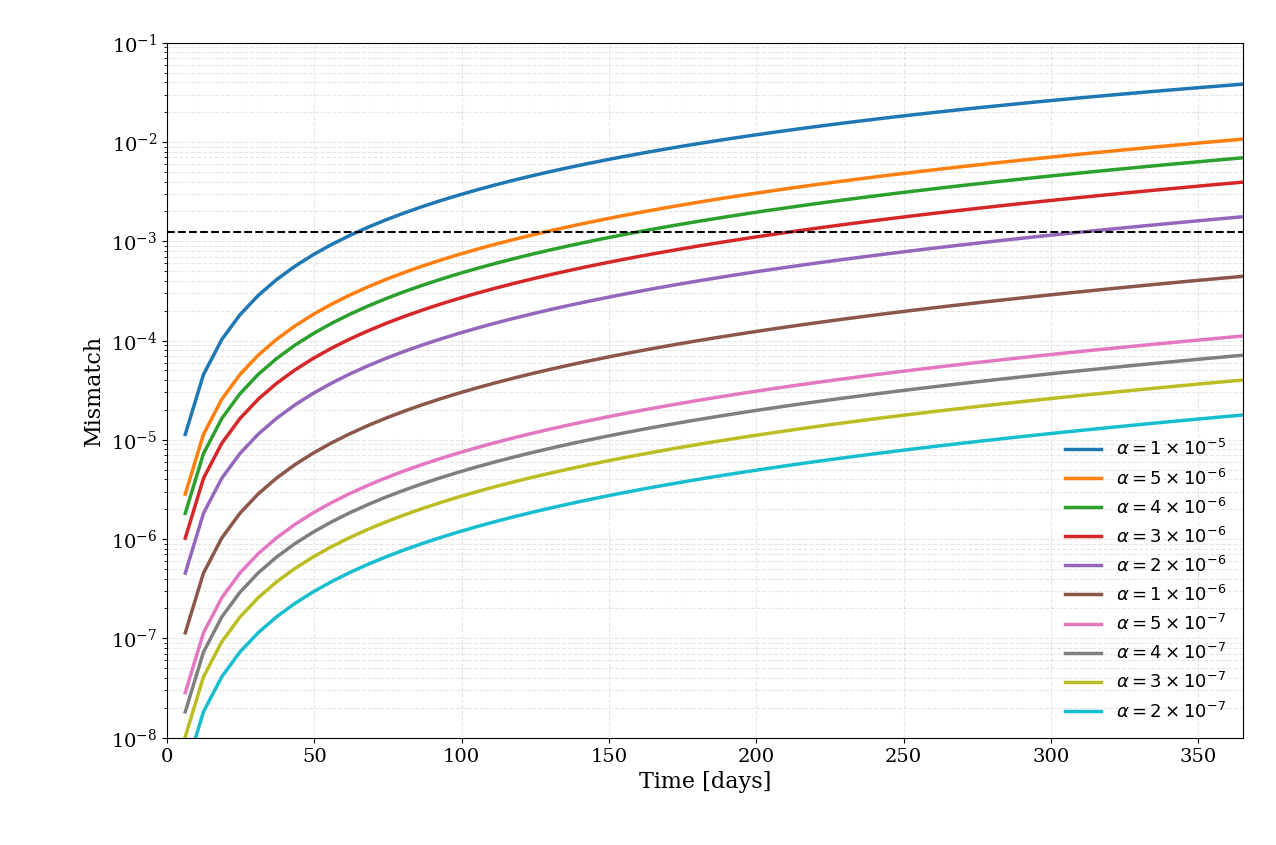}
    \caption{The plot illustrates the dependence of the mismatch on the observation time for the different values of the CoS parameter.}
    \label{fig:mismatch}
\end{figure*}

\section{Gravitational waveforms of EMRI system and data analysis}\label{sec:GWforms}

In this section, we investigate the gravitational waveforms of the EMRI system. One can write the metric perturbations that describe the GWs for the EMRI system as~\cite{RevModPhys.52.299}
\begin{equation}
h_{ij}=\frac{2}{D_L}\ddot{Q}_{ij}\ ,
\end{equation}
where $D_L$ represents the luminosity distance between the EMRI system and the detector. To measure the gravitational-wave polarizations, we must first set up a coordinate system based on the detector's orientation $(X,Y,Z)$~\cite{Poisson_Will_2014}. After that, we can write the GW polarizations as follows~\cite{PhysRevD.75.024005,Yang:2024cnd}
\begin{eqnarray}
h_{+}&=&\frac{1}{2}(h_{\zeta\zeta}-h_{\iota\iota}), \\
h_{\times}&=& h_{\iota\zeta}.
\end{eqnarray}
with
\begin{widetext}
\begin{eqnarray}
h_{\zeta\zeta}&=&h_{xx}\cos^2{\zeta}-h_{xy}\sin{2\zeta}+h_{yy}\sin^2{\zeta}, \\
h_{\iota\iota}&=&\cos{\iota}[h_{xx}\sin^2{\zeta}+h_{xy}\sin{2\zeta}+h_{yy}\cos^2{\zeta}]+h_{zz}\sin^2{\iota}-\sin{2\iota}[h_{xz}\sin{\zeta}+h_{yz}\cos{\zeta}], \\
h_{\iota\zeta}&=&\cos{\iota}\Big[\frac{1}{2}h_{xx}\sin{2\zeta}+h_{xy}\cos{2\zeta}-\frac{1}{2}h_{yy}\sin{2\zeta}\Big]+\sin{\iota}[h_{yz}\sin{\zeta}-h_{xz}\cos{\zeta}].
\end{eqnarray}
\end{widetext}
where $\zeta$ and $\iota$ are the latitude and the angle of inclination, respectively. In this article, we set them as $\zeta=\iota=\pi/4$. We can then rewrite the above equation in the following form
\begin{eqnarray}
h_{+}&=&\frac{1}{4D_L}(\ddot{Q}_{xx}+\ddot{Q}_{yy})-\frac{3}{2D_L}\ddot{Q}_{xy}-\frac{1}{2D_L}\ddot{Q}_{zz} \ , \nonumber \\
h_{\times}&=& \frac{\sqrt{2}}{2D_L}(\ddot{Q}_{xx}-\ddot{Q}_{yy})\ .
\end{eqnarray}
We consider the luminosity distance as $D_L=2Gpc$. In Fig.~\ref{fig:GW}, we plot the gravitational waveforms for the different values of the CoS parameter. The figure shows that initially, the impact of the CoS parameter on the waveforms is negligible. However, after one year, the dephasing caused by the CoS becomes clearly distinguishable. 

We can assess the detectability of the effect of the CoS parameter by analyzing the GW signal as measured by a detector. The characteristics of the detector determine how it responds to GW. The strain amplitude can be written in LISA in the following form~\cite{Cutler:1997ta}
\begin{equation}
h_{I,II}=\frac{\sqrt{3}}{2}(F_{I,II}^{+}h_{+}+F_{I,II}^{\times}h_{\times})\,,
\end{equation}
where $\sqrt{3}/2$ refers to the actual angle between the LISA arms, which is 60$^\circ$, and $F_{I,II}^{+}$ and $F_{I,II}^{\times}$ denote the antenna pattern functions~\cite{Barack:2003fp}. These functions depend on the directions of the source and orbital angular momentum, which are $(\theta_S,\phi_S)$ and $(\theta_L=0,\phi_L=0)$, respectively. 
To quantify the distinguishability between waveforms, we calculate the mismatch of the ordinary Schwarzschild BH and the Schwarzschild BH surrounded by a CoS. In matched filtering, the accuracy of parameter estimation depends on the degree to which the template waveform $h_T$ aligns with the actual signal $h_S$. The mismatch function quantifies any discrepancy between them; therefore, a minimal mismatch is essential for reliable results. It can be written as follows~\cite{PhysRevD.78.124020,Cutler:1997ta}
\begin{equation}
\mathcal{M}(h_T,h_S)=1-\frac{\langle h_T|h_S\rangle}{\sqrt{\langle h_T|h_T\rangle\langle h_S|h_S\rangle}}\ ,
\end{equation}
with
\begin{equation}
\langle h_T|h_S\rangle=4 \text{Re}\int_0^{\infty} \frac{\tilde{h}_{T}(f)\tilde{h}_S^{*}(f)}{S_n(f)}df\,,
\end{equation}
where $\tilde{h}(f)$ is the Fourier transform of the waveforms and $S_n(f)$ is the detector's one-sided noise power spectral density. In Fig.~\ref{fig:mismatch}, we demonstrate the mismatch between the ordinary Schwarzschild BH and the Schwarzschild BH surrounded by a CoS as a function of the observation time (one year for this figure) for the different values of the CoS parameter. The figure shows that the mismatch increases with both the observation time and the value of $\alpha$. A common threshold for detectability in GW data analysis is a mismatch of $\mathcal{M} \simeq 0.00125$ (see Refs.~\cite{zi2025,Xia:2025yzg}). Using this threshold, Fig.~\ref{fig:mismatch} indicates that LISA could detect the presence of a CoS for the parameter values $\alpha \gtrsim 2 \times 10^{-6}$. The effect of the smaller values of the CoS parameter than $\alpha \approx 2 \times 10^{-6}$ on the gravitational waveforms remains indistinguishable by LISA during the year of observation.

\section{Conclusions}\label{summary}

In this work, we investigate the quasi-circular EMRIs by considering the supermassive Schwarzschild BH surrounded by CoS. First, we reviewed the spacetime geometry of the Schwarzschild BH in the presence of CoS. After that, we explored the motion of a test particle around a Schwarzschild BH in the presence of CoS using the Lagrangian formalism. In particular, we investigated the effective potential for a massive particle around a Schwarzschild BH surrounded by CoS. We found that the values of the effective potential decrease under the influence of the CoS parameter and vice versa for the orbital angular momentum (see Fig.~\ref{fig:eff}). Moreover, we explore the effect of the CoS parameter on the MBO and ISCO.  
We plot the radial dependence of $\dot{r}^2$ for the different values of the energy of the test particle with the fixed CoS parameter in Fig.~\ref{fig:rdot}. For bound orbits, there are two relevant points where $\dot{r}^2=0$, corresponding to periapsis ($r_p$) and apoapsis ($r_a$). Subsequently, we consider the eccentric motion, and we find the energy and angular momentum of the test particles by considering the Euler-Lagrange equations.

In addition, we investigated the effect of CoS on the dynamics of the EMRI system. Specifically, the effect of the CoS parameter on the orbital evolution of the semi-latus rectum and eccentricity was demonstrated in Fig.~\ref{fig:pe}. To provide more information, we plot the quasi-circular orbits of the secondary object around the supermassive Schwarzschild BH surrounded by a CoS in Fig.~\ref{fig:trac}. 

Furthermore, the effect of the CoS parameter on the gravitational waveforms of the EMRI system was studied. Our analysis of the gravitational waveforms (Fig.~\ref{fig:GW}) revealed that while the initial waveform is largely insensitive to $\alpha$, a significant dephasing accumulates over a one-year observation period. We then turn to the LISA data analysis. We explore the mismatch between the ordinary Schwarzschild BH and the Schwarzschild BH surrounded by a CoS. The dependence of the mismatch on the observation time is illustrated in Fig.~\ref{fig:mismatch}. Based on our mismatch analysis, we conclude that LISA has the potential to distinguish a Schwarzschild BH with a CoS (when $\alpha \gtrsim 2 \times 10^{-6}$) from a standard Schwarzschild BH.

It should be emphasized that this is a primary investigation based on the flat-spacetime quadrupole equation. Therefore, we aim to perform a more precise analysis using the perturbation theory in the future.

\section*{Acknowledgements}

MA warmly thanks Qiyuan Pan for their valuable comments and discussions on this topic. This research was funded by the National Natural Science Foundation of China (NSFC) under Grant No. U2541210.

\bibliographystyle{apsrev4-1}
\bibliography{ref}

\end{document}